\documentclass[aps,floats,floatfix,twocolumn,amsmath,amssymb]{revtex4-1}

\usepackage{sidecap} 

\usepackage{graphicx}
\usepackage{dcolumn}
\usepackage{bm}

\begin{document}

\title{Interaction of a point charge with the surface of a uniaxial dielectric}

\author{Primo\v{z} Rebernik Ribi\v{c}}
\affiliation{Facult\'{e} des Sciences de Base, Ecole Polytechnique F\'{e}d\'{e}rale de Lausanne (EPFL), Lausanne, Switzerland}

\author{Rudolf Podgornik}
\affiliation{Department of Theoretical Physics, J. Stefan Institute and Department of Physics, Faculty of Mathematics and Physics, University of Ljubljana, SI-1000 Ljubljana, Slovenia}

\date{\today}

\begin{abstract}
We analyze the force on a point charge moving at relativistic speeds parallel to the surface of a uniaxial dielectric. Two cases are examined: a lossless dielectric with no dispersion and a dielectric with a plasma type response. The treatment focuses on the peculiarities of the strength and direction of the interaction force as compared to the isotropic case. We show that a plasma type dielectric can, under specific conditions, repel the point charge.
\end{abstract}


\maketitle

\section{Introduction}
Despite the long and rich history of theoretical studies on the interaction between fast charges and solid surfaces (see, e.g. refs. \cite{morozov, bolotovskii, takimoto, mills:1977, muscat, barberan, mahanty, zutter, mills:1992, schieber:1998, schieber:2000, schieber:2002}) unexpected results can be and indeed are derived. A recent example is the discovery by one of the authors of the present letter \cite{ribic}, that the interaction between a relativistic charge packet and a metal or dielectric surface can become repulsive by simply tuning the packet geometry; a result that seems to go against common notions established in electrodynamics and should be of importance in the framework of accelerator physics and electron spectroscopy.

In this letter we switch gears and focus on the interaction between a point charge and a uniaxial dielectric within the context of ionic and molecular interactions with macroscopic surfaces. We assume a description of the surface that approximates the non-isotropic nature of crystalline surfaces and surfaces decorated with adsorbed non-isotropic inclusions. We present a derivation of the force on the charged particle starting from Maxwell's equations in Fourier space and then evaluate the force numerically in real space. We show that the longitudinal component of the force (parallel to the dielectric surface) is in general not parallel to the particle velocity anymore and that its direction depends on the particle speed (energy). We demonstrate two peculiarities of the plasma type response: 1) the direction of the longitudinal force depends on the distance of the particle from the surface and 2) under specific conditions the particle can be repelled by the surface.

\section{Evaluation of the electromagnetic force}
\begin{figure}
\includegraphics[width=0.4\textwidth]{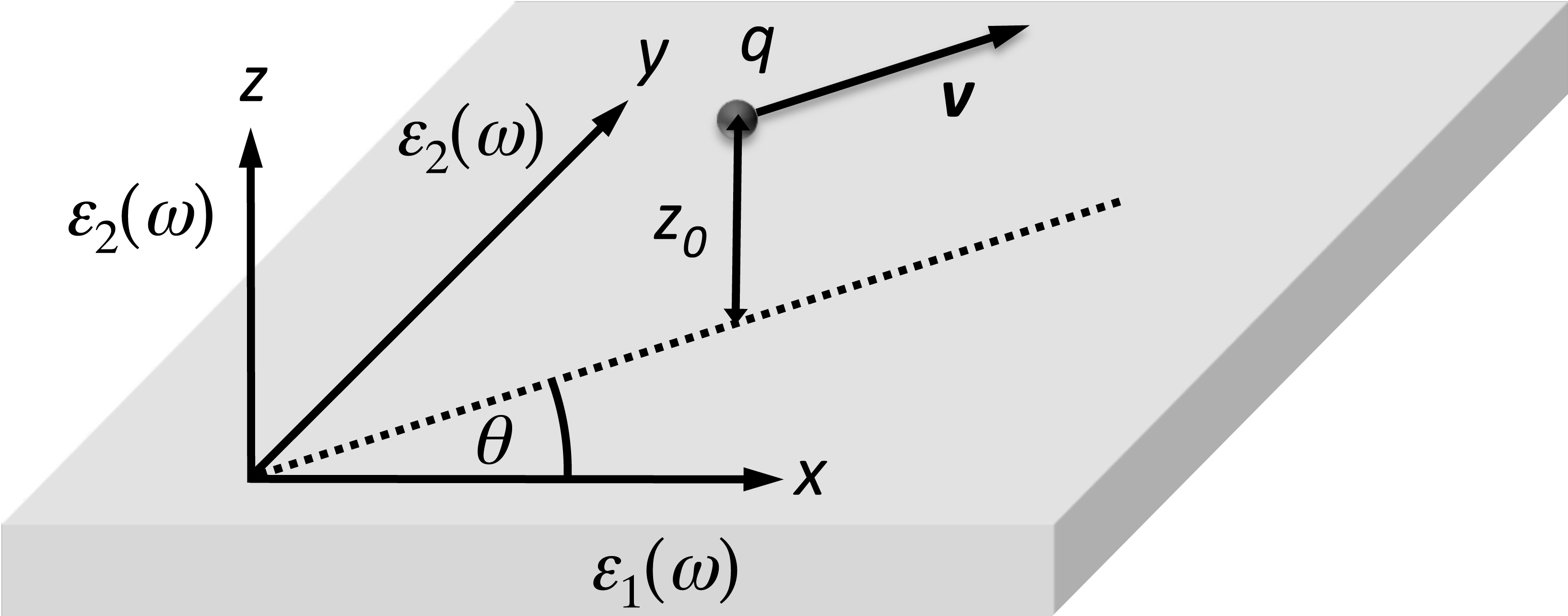}
\caption{The point charge moves in vacuum at a distance $z_0$ parallel to the surface of a uniaxial dielectric defined by the $xy$ plane. The velocity makes an angle $\theta$ with the optical axis ($x$-direction). The dielectric tensor is diagonal with $\epsilon_{xx}=\epsilon_1 (\omega)$ and $\epsilon_{yy}=\epsilon_{zz}=\epsilon_2 (\omega)$.}
\label{fig.1}
\end{figure}

The geometry of the problem is illustrated in fig.~\ref{fig.1}. A point charge moves in vacuum with a velocity $\mathbf{v}$ at a distance $z_0$ parallel to the surface of a uniaxial dielectric. The dielectric surface lies in the $xy$ plane with the optical axis oriented along the $x$ direction. The velocity is $\mathbf{v}=(v_x,v_y,0)=(v\cos{\theta},v\sin{\theta},0)$, where $\theta$ is the angle between $\mathbf{v}$ and the optical axis.

The electromagnetic field due to the moving point charge is calculated from Maxwell's equations:
\begin{align}
\nabla \cdot \mathbf{D} &=  \rho \label{Maxwell1} \\
\nabla \times \mathbf{H} &= \mathbf{J} + \frac{\partial \mathbf{D}}{\partial t} \mbox{.}
\label{Maxwell2}
\end{align}

The problem is tackled by replacing the fields with the standard scalar and vector potentials defined as $\mathbf{B}=\nabla \times \mathbf{A}$ and $\mathbf{E}=-\nabla \Phi - \partial \mathbf{A}/ \partial t$. Then the solution to eqs.~(\ref{Maxwell1}) and (\ref{Maxwell2}) is sought separately in the vacuum ($z>0$) and dielectric ($z<0$) half spaces by introducing three-dimensional Fourier transforms of all quantities:
\begin{equation}
G(\mathbf{r},z,t)=\int \frac{d^2\mathbf{k} d\omega}{(2\pi)^3}g(\mathbf{k}, z,\omega) e^{i(\mathbf{k} \cdot \mathbf{r}-\omega t)} \mbox{,}
\label{Fourier}
\end{equation}
where $\mathbf{k}=(k_x,k_y)$ and $\mathbf{r}=(x,y)$ are the wave and position vectors parallel to the dielectric interface.

The absence of bound charges and currents in vacuum allows us to decouple the above equations using the Lorenz gauge:
\begin{align}
\frac{\partial^2 \Phi_1}{\partial z^2}-Q_1^2\Phi_1&=-\frac{\rho}{\epsilon_0}  \label{vacuum1} \\
\frac{\partial^2 \mathbf{A}_1}{\partial z^2}-Q_1^2\mathbf{A}_1&=-\frac{\mathbf{J}}{c^2\epsilon_0}  \label{vacuum2} \mbox{,}
\end{align}
where $Q_1=\sqrt{k^2-\omega^2/c^2}$, $k=\sqrt{k_x^2+k_y^2}$ is the magnitude of $\mathbf{k}$ and the index 1 refers to the vacuum half space. The Fourier decomposition of the charge density is:
\begin{equation}
\rho (\mathbf{k},z,\omega)=2 \pi q \delta(\omega-\mathbf{v} \cdot \mathbf{k}) \delta(z-z_0) \mbox{,}
\label{rho}
\end{equation}
while the current density is simply $\mathbf{J}= \rho \mathbf{v}$.

The solution to eqs.~(\ref{vacuum1}) and (\ref{vacuum2}) is written as a sum of the "incident" field due to the point charge (the same as in the absence of the dielectric interface):
\begin{align}
\Phi_i &= \frac{\pi q}{\epsilon_0}\delta(\omega - \mathbf{v} \cdot \mathbf{k}) \frac{e^{-Q_1 \lvert z-z_0\rvert }}{Q_1} \label{incidentphi} \\
\mathbf{A}_i &= \frac{\mathbf{v}}{c^2} \Phi_i  \mbox{,}
\label{incidenta}
\end{align}
and the "scattered" field due to the dielectric:
\begin{equation}
\mathbf{A}_s = \mathbf{a}_s e^{-Q_1 z}  \mbox{,}
\label{scattered}
\end{equation}
where for the latter we impose the gauge $\Phi_s=0$ and therefore $\nabla \cdot \mathbf{A}_s=0$.

In these equations $Q_1$ is a real number. This follows from the fact that all the fields carry the prefactor $\delta(\omega - \mathbf{v} \cdot \mathbf{k})$, which insures that $\omega = \mathbf{v} \cdot \mathbf{k}$. Writing $\mathbf{k}=(k \cos{\phi}, k \sin{\phi})$ we obtain $Q_1 = k \sqrt{1 - \beta^2 \cos^2{(\theta - \phi)}}$, a real quantity for any $\beta=v/c$, $\theta$ and $\phi$. The point charge moving at constant speed parallel to the dielectric surface cannot emit radiation in the vacuum half space - the waves in eqs.~(\ref{incidentphi}) - (\ref{scattered}) are evanescent and the sign in front of $Q_1$ in eq.~(\ref{scattered}) is negative to insure that the waves decay to zero as $z\rightarrow \infty$ 

The following linear constitutive relations are assumed for the dielectric: $\mathbf{D}=\epsilon_0 \boldsymbol\epsilon \cdot \mathbf{E}$, where $\boldsymbol\epsilon$ is diagonal with $\epsilon_{xx}=\epsilon_1 (\omega)$ and $\epsilon_{yy}=\epsilon_{zz}=\epsilon_2 (\omega)$ and $\mathbf{H}=\mathbf{B}/\mu_0$ (the material is non-magnetic). From eq.~(\ref{Maxwell1}) we can again set $\Phi_2=0$ and therefore $\nabla \cdot (\boldsymbol\epsilon \cdot \mathbf{A}_2)=0$. The equation for the vector potential in the dielectric obtained by taking the curl of eq.~(\ref{Maxwell2}) then becomes:
\begin{align}
\frac{\partial^2 A_{2x}}{\partial z^2}-Q_{2e}^2 A_{2x} &=0 \nonumber \\
\frac{\partial^2 A_{2y}}{\partial z^2}-Q_{2o}^2 A_{2y} &=\left( \frac{\epsilon_1}{\epsilon_2} -1 \right)k_x k_y A_{2x} \nonumber \\
\frac{\partial^2 A_{2z}}{\partial z^2}-Q_{2o}^2 A_{2z} &=\left( \frac{\epsilon_1}{\epsilon_2} -1 \right)(-i k_x) \frac{\partial A_{2x}}{\partial z} \mbox{,} \label{dielectric}
\end{align}
where $Q_{2o}=\sqrt{k^2-\epsilon_2 (\omega^2/c^2)}=k\sqrt{1-\epsilon_2 \beta^2 \cos^2{(\theta - \phi)}}$ and $Q_{2e}=\sqrt{(\epsilon_1/ \epsilon_2) k_x^2 + k_y^2 -\epsilon_1 (\omega^2/c^2)}$ $=$ $k\sqrt{\frac{\epsilon_1}{\epsilon_2} \cos^2{\phi} + \sin^2{\phi-\epsilon_1 \beta^2 \cos^2{(\theta-\phi)}}}$, where index 2 refers to the dielectric half space. The above equations are consistent with results obtained by other authors (see, e.g.~\cite{barash,fleck}), except that we prefer to work with potentials rather than fields.

The general solution to eq.~(\ref{dielectric}) is written as a sum of ordinary (o) and extraordinary (e) waves:
\begin{equation}
\mathbf{A}_2=\mathbf{a}_o e^{Q_{2o} z} + \mathbf{a}_e e^{Q_{2e} z} \mbox{,}
\label{soldielectric}
\end{equation}
where
\begin{align}
\mathbf{a}_o&=(0,a_{oy},a_{oz}) \nonumber \\
\mathbf{a}_e&=a_e\left (1,\frac{k_x k_y}{k_x^2-\epsilon_2 \frac{\omega^2}{c^2}},\frac{- i k_x Q_{2e}}{k_x^2-\epsilon_2 \frac{\omega^2}{c^2}} \right ) \label{soldielectricsup} \mbox{.}
\end{align}

Since both $\epsilon_1(\omega)$ and $\epsilon_2(\omega)$ are in general complex quantities, $Q_{2o}$ and $Q_{2e}$ are also complex. There are two solutions for $Q_{2o}$ and $Q_{2e}$ but the physical ones correspond to those with positive real parts (only these decay exponentially in the dielectric and satisfy the radiation condition).

To find the coefficients contained in the vectors $\mathbf{a}_s$, $\mathbf{a}_o$ and $\mathbf{a}_e$ we impose the usual boundary conditions for the fields at the interface. The procedure, although straightforward, is tedious and will not be reproduced in detail. Using the obtained coefficients the Lorentz force components are:
\begin{align}
f_x&=i q e^{-Q_1 z_0}  \left[\omega a_{sx} + v_y (k_x a_{sy} - k_y a_{sx})\right] \nonumber \\
f_y&=i q e^{-Q_1 z_0}  \left[\omega a_{sy} + v_x (k_y a_{sx} - k_x a_{sy})\right] \nonumber \\
f_z&=-q Q_1 e^{-Q_1z_0} (v_x a_{sx} + v_y a_{sy}) \mbox{,}
\label{force}
\end{align}
which become, after the Fourier transform over $\omega$ 
:
\begin{align}
f_x&=-i \frac{k_x}{Q_1}f_z \nonumber \\
f_y&=-i \frac{k_y}{Q_1}f_z \nonumber \\
f_z&=\frac{e^{-2 Q_1 z_0} q^2}{2 c^2 \epsilon_0 } \frac{R}{P}  \mbox{,}
\label{forcesolution}
\end{align}
where $R$ and $P$ are defined as:
\begin{widetext}
\begin{eqnarray}
R &=& -\epsilon_2(\mathbf{v} \cdot \mathbf{k})^2 \bigg\{v_x^2 Q_{2o} (Q_{2e}-Q_1) (Q_{2o}+Q_1) + v_y^2 Q_{2o} (Q_{2o}-Q_1) (Q_{2e}+Q_1) +2 (\epsilon_2-1) k_x k_y v_x v_yQ_1  \nonumber \\ &+& (\epsilon_2-1) k_y^2 \Big[v_x^2(Q_{2e}-Q_1) + v_y^2 (Q_{2e}+Q_1) \Big] \bigg\} +c^4 k_x^2 Q_{2o} (Q_{2o}+Q_1) (Q_1- \epsilon_2 Q_{2e})\nonumber \\&+&c^2 \Bigg\langle 2(\epsilon_2 - 1) k_x^3 k_y v_x v_y Q_1  + \epsilon_2 k_y^2 v_y^2 (Q_{2e}+Q_1) \Big[Q_{2o} (Q_{2o}-Q_1)+(\epsilon_2 - 1) k_y^2 \Big] \nonumber \\&+&k_x^2 \bigg\{Q_1 Q_{2o}  \Big[-v_x^2(Q_{2o}+Q_1) +v_y^2(Q_{2o}-Q_1) \Big]+\epsilon_2 Q_{2o} \Big[v_x^2(2 Q_{2e}-Q_1) (Q_{2o}+Q_1) +v_y^2 Q_{2e} (Q_{2o}-Q_1) \Big] \nonumber \\&+&(\epsilon_2 - 1)k_y^2  \Big[2\left(-v_x^2+v_y^2\right) Q_1 + \epsilon_2 v_x^2 (Q_{2e}+Q_1) \Big]\bigg\}+2 k_x k_y v_x v_y \bigg\{(\epsilon_2 - 1) k_y^2  \Big[(\epsilon_2 - 1) Q_1 \nonumber \\&+& \epsilon_2 Q_{2e} \Big]+Q_{2o} \Big[\epsilon_2 Q_1 (Q_{2e}-Q_1) +Q_{2o} (\epsilon_2 Q_{2e} - Q_1)\Big]\bigg\}\Bigg\rangle
\label{R}
\end{eqnarray}
\begin{equation}
P= c^2 k_x^2 Q_{2o} (Q_1+Q_{2o}) (Q_1+\epsilon_2Q_{2e} ) -\epsilon_2 (Q_1+Q_{2e}) (\mathbf{v} \cdot \mathbf{k})^2 \left[Q_{2o} (Q_1+Q_{2o})+k_y^2 (\epsilon_2 - 1)\right] 
\label{P}
\end{equation}
\end{widetext}

The force in real space is obtained by integration of the above expressions over $\mathbf{k}=(k \cos{\phi}, k \sin{\phi})$, setting $\mathbf{r}=\mathbf{v}t$. For a lossless and dispersionless material the transform over $k$ is performed analytically, giving an inverse second power dependence of the force on the distance $z_0$ from the surface, while the transform over $\phi$ has to be performed numerically. When dispersion and/or losses are included the integration can be performed only numerically.

The longitudinal force $\mathbf{f}_p=(f_{x},f_{y})$ can be interpreted in Fourier space in a convenient way. The point particle excites electromagnetic waves in the semiinfinite dielectric and each of these waves carries a momentum proportional to $\mathbf{k}$. This momentum has to be balanced by the particle which results in a force parallel to the interface. The magnitude of the momentum is determined by the boundary conditions and material properties. The longitudinal force $\mathbf{F}_p=(F_{x},F_{y})$ in real space is then obtained by integration over the momenta of all the excited waves.

In the following we treat two examples of dielectric response: a lossless dielectric with no dispersion and a plasma type response. Mathematically both cases can be analyzed using the Drude model for the dielectric tensor:
\begin{align}
\epsilon_1(\omega)&=\epsilon_{\infty1}-\frac{\omega_{p1}^2}{\omega (\omega + i \gamma_1)} \nonumber \\
\epsilon_2(\omega)&=\epsilon_{\infty2}-\frac{\omega_{p2}^2}{\omega (\omega + i \gamma_2)} \mbox{,}
\end{align}
where $\epsilon_{\infty1,\infty2}$ are dielectric constants at high frequencies ($\omega \rightarrow \infty$), $\omega_{p1,p2}$ are the plasma frequencies and $\gamma_{1,2}$ are the damping coefficients. For a lossless dielectric $\omega_{p1}$, $\omega_{p2}$ and $\gamma_{1}$, $\gamma_{2}$ are made infinitely small, i.e. $\epsilon_{1,2}(\omega)=\epsilon_{\infty1,\infty2}$, while for a plasma type dielectric we take $\epsilon_{\infty1}=\epsilon_{\infty2}=1$. 
For reasons of consistency the frequency has to satisfy:
\begin{equation}
\omega \gg \gamma_{1,2} \mbox{.}
\label{plasmacondition}
\end{equation}

It can be shown (see, e.g. ~\cite{jackson}) that the range of frequencies a point charge moving above a solid excites near its surface is proportional to $\gamma v/z_0$, where $\gamma=(1-v^2/c^2)^{-1/2}$ is the relativistic factor. The condition of eq.~(\ref{plasmacondition}) therefore reads:
\begin{equation}
\omega_{max}=\frac{\gamma v}{z_0} \gg \gamma_{1,2}  \mbox{.}
\end{equation}

In addition to the above, for non-zero losses the plasma model is not applicable near $\omega_{p1,p2}$, where the imaginary part of the dielectric function dominates the response.

For a lossless dielectric with no dispersion, i.e. $\epsilon_1(\omega)=\epsilon_{\infty1} $ and $\epsilon_2(\omega)=\epsilon_{\infty2} $ are constants, $\mathbf{F}_p$ will be non-zero only if the \v{C}erenkov condition, either for the ordinary or extraordinary waves (or both), is satisfied. This occurs when either $Q_{2o}$ or $Q_{2e}$ becomes imaginary, i.e. the waves become propagating. If both $Q_{2o}$ and $Q_{2e}$ are real, the waves excited in the solid are evanescent (they decay exponentially in the solid) and these do not contribute to $\mathbf{F}_p$. The reason is that $\mathbf{F}_p$ in general decreases the energy of the particle and this can occur only if the charge emits radiation into the dielectric.

It is straightforward to show that $Q_{2o}$ becomes imaginary if the particle moves faster than the critical speed:
\begin{equation}
\beta_o=v_o/c=\frac{1}{\sqrt{\epsilon_2}}  \mbox{.}
\label{cerenkov_ord}
\end{equation}
Above $\beta_o$ propagating \v{C}erenkov waves are emitted into a circular cone defined by:
\begin{equation}
\cos{\alpha_o}=\frac{1}{\beta \sqrt{\epsilon_2}}  \mbox{.}
\label{cerenkov_ord_angle}
\end{equation}
Here $\alpha_o$ is the angle between the optical axis and the wave vector $\mathbf{K}=(k_x, k_y, \mbox{sgn}[\omega] k_z)$, and $k_z=i Q_{2o}$. The function $\mbox{sgn}{(\omega)} $ insures that the energy flow is directed into the dielectric (propagating waves are actually emitted only into one half of the \v{C}erenkov cone). The symmetry axis of the ordinary \v{C}erenkov cone is always parallel to the particle velocity.

For extraordinary waves the analysis is a bit more cumbersome \cite{delbart}. The critical speed for \v{C}erenkov emission is:
\begin{equation}
\beta_e=\frac{1}{\sqrt{\epsilon_1 \sin^2{\theta} + \epsilon_2 \cos^2{\theta}}}  \mbox{.}
\label{cerenkov_ext}
\end{equation}
From eq.~(\ref{cerenkov_ext}) it follows that for given $\epsilon_1$, $\epsilon_2$, and $\beta$ there exists a critical angle $\theta_c$ above which the \v{C}erenkov condition for extraordinary waves is satisfied:
\begin{equation}
\cos{\theta_c}= \frac{1}{\beta}\sqrt{\frac{\beta^2 \epsilon_1 -1}{\epsilon_1 - \epsilon_2}} \mbox{.}
\label{critical_angle_cerenkov_ext}
\end{equation}

The extraordinary \v{C}erenkov cone has an elliptical cross section \cite{delbart}, which follows from the definition of $Q_{2e}$. The symmetry axis of the cone lies in the $xy$ plane and makes an angle $\chi$ with the $x$ axis \footnote{To obtain $\chi$ we write the equation for $Q_{2e}$ as $\mathbf{K} \cdot (\mathbf{M} \cdot \mathbf{K}) = 0$, where $\mathbf{K}=(k_x, k_y, k_z)$, $k_z=i Q_{2e}$ and $\mathbf{M}$ is a $3 \times 3$ matrix. In the coordinate system where $\mathbf{M}$ is diagonal the equation for $Q_{2e}$ defines the extraordinary \v{C}erenkov cone. The transformation to this coordinate system involves a rotation by $\chi$ along the $z$ axis. The result reduces to the one for ordinary waves when $\epsilon_1=\epsilon_2$.}:
\begin{widetext}
\begin{equation}
\tan{\chi} = \frac{\epsilon_1 - \epsilon_2 + \sqrt{(\epsilon_1 - \epsilon_2)^2 + (\beta^2 \epsilon_1 \epsilon_2)^2 + 2 \beta^2 \epsilon_1 \epsilon_2 (\epsilon_2 - \epsilon_1)\cos{(2 \theta)} }}{\beta^2 \epsilon_1 \epsilon_2 \sin{(2 \theta)}} -\cot{(2 \theta)}  \mbox{,}
\label{cerenkov_sym_ext}
\end{equation}
\end{widetext}
which for $\beta=\beta_e$ and $\theta=\theta_c$ reduces to:
\begin{equation}
\tan{\chi} = \frac{\epsilon_1}{\epsilon_2} \tan{\theta_c} \mbox{.}
\label{cerenkov_sym_ext_crit}
\end{equation}

For a plasma type dielectric $\mathbf{F}_p$ is in general always nonzero, but is appreciable only in a certain frequency range due to condition (\ref{plasmacondition}). This can be seen for the limiting case when $\gamma_1=\gamma_2=0$ and therefore $\epsilon_{1,2}(\omega)=1-\omega^2_{p1,2}/\omega^2$. If follows that $Q_{2o}$ is real for all $\omega$ and there are no ordinary propagating waves. On the other hand $Q_{2e}$ is imaginary if the frequency $\omega$ lies between the plasma frequencies \cite{galyamin}: e.g., $\omega_{p1} < \omega < \omega_{p2}$ for $\omega_{p2} > \omega_{p1}$. We will refer to this frequency range the \v{C}erenkov band. For zero losses, only waves in the \v{C}erenkov band contribute to the longitudinal force, $\mathbf{F}_p$.

\section{Results}
\begin{figure}
\includegraphics[width=0.4\textwidth]{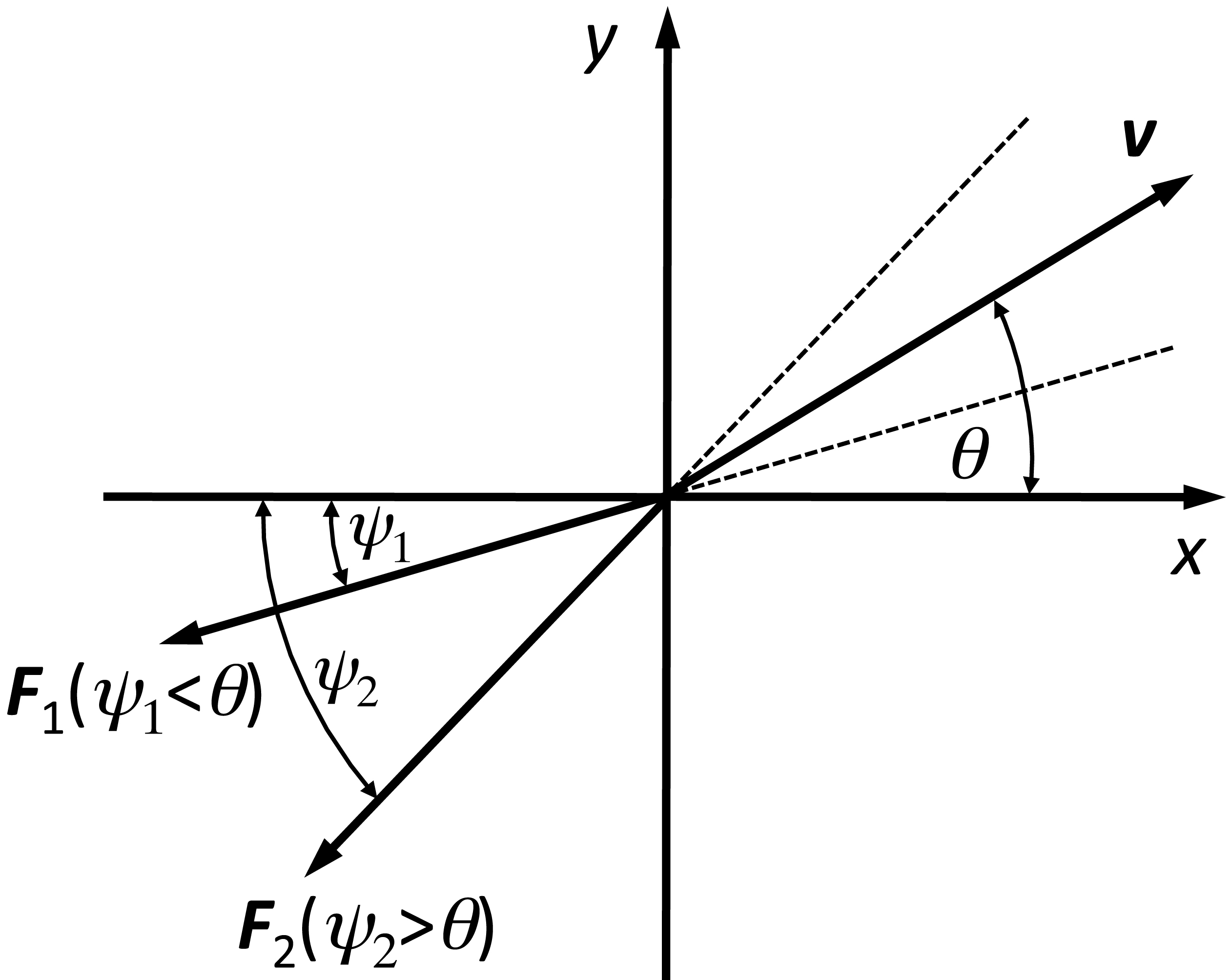}
\caption{Different scenarios for the direction of the longitudinal component of the Lorentz force acting on the electron.}
\label{fig.2}
\end{figure}

In this section we plot the magnitude of $\mathbf{F}_p$, the angle between $\mathbf{F}_p$ and the electron velocity, and the transverse force $F_z$. For an isotropic material, where $\epsilon_1(\omega)=\epsilon_2(\omega)$, $\mathbf{F}_p$ is in the direction opposite to $\mathbf{v}$, while for a uniaxial material this is no longer the case. We plot $\psi=\arctan{(F_{y}/F_{x})}$ as a function of $\theta$ to show the effect of the anisotropy (the angle between $\mathbf{v}$ and $\mathbf{F}_p$ is equal to $\psi + \pi$). Figure~\ref{fig.2} sketches two possible scenarios that may occur.

\begin{figure*}
\includegraphics[width=0.7\textwidth]{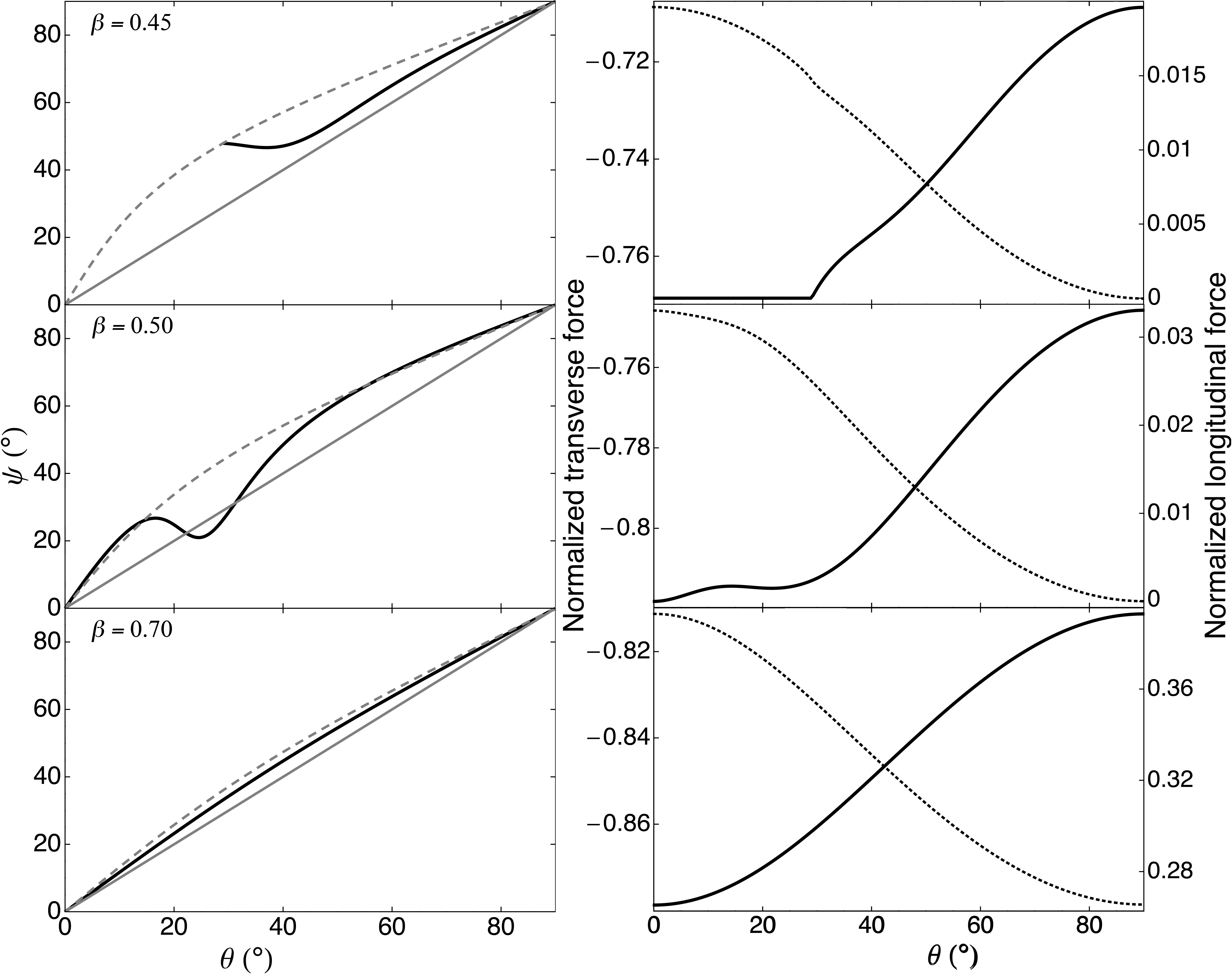}
\caption{Results for a dispersionless and lossless dielectric with $\epsilon_1=8$ and $\epsilon_2=4$. Left column: $\psi=\arctan{(F_{y}/F_{x})}$ as a function of $\theta$ for $\beta=0.45$ (top), $\beta=0.5$ (middle) and $\beta=0.7$ (bottom). Solid black line represents the results, solid gray line is the isotropic case ($\psi=\theta$, valid for $\beta$ above the \v{C}erenkov condition) and dashed gray line is the direction $\chi$ of the symmetry axis of the extraordinary \v{C}erenkov cone. Right column: magnitude of the longitudinal force (solid line) and transverse force (dashed line) as a function of $\theta$ for $\beta=0.45$ (top), $\beta=0.5$ (middle) and $\beta=0.7$ (bottom).}
\label{fig.3}
\end{figure*}

In fig.~\ref{fig.3} we show the results for a dispersionless material with no losses. We chose $\epsilon_1=8$ and $\epsilon_2=4$. In the left column we plot $\psi$ as a function of $\theta$ for different values of $\beta=v/c$, while in the right column we plot the magnitude of the longitudinal force and the transverse force, both normalized with respect to $q^2/(16 \pi \epsilon_0 z_0^2)$ (static image charge force for a charge above a metal surface).

\begin{figure*}
\includegraphics[width=0.7\textwidth]{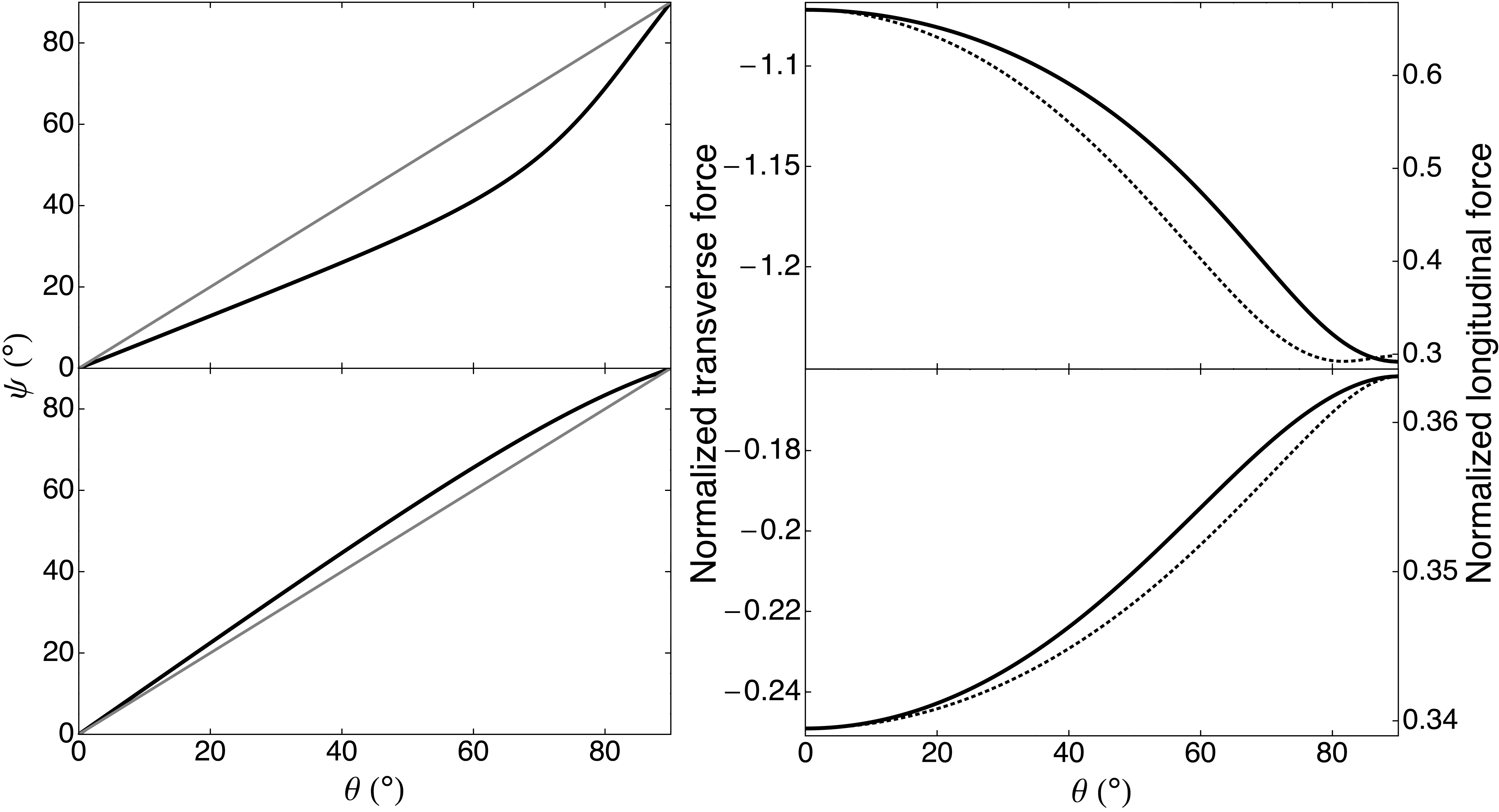}
\caption{Results for a plasma type dielectric ($\gamma_{1,2}=10^{-2}\omega_{p1}$): $\psi=\arctan{(F_{y}/F_{x})}$ as a function of $\theta$ (left column) and force components (right column). Top panels:  $\omega_{p2}=3\omega_{max}$, $\omega_{p1}=10^{-1}\omega_{p2}$.  Bottom panels:  $\omega_{p2}=0.5\omega_{max}$, $\omega_{p1}=10^{-1}\omega_{p2}$. Solid black line in the left column represents the results, while the solid gray line is the isotropic case ($\psi=\theta$). Solid line in the right column is the magnitude of the longitudinal force, while dotted line represents the transverse force.}
\label{fig.4}
\end{figure*}

The \v{C}erenkov condition for ordinary waves, $\beta_o >0.5$, is only satisfied for bottom panels in fig.~\ref{fig.3}. For extraordinary waves eq.~(\ref{critical_angle_cerenkov_ext}) gives $\theta_c \approx 29^{\circ}$. From fig.~\ref{fig.3}, $F_p$ is zero below this value and increases with $\theta$. The direction of $\mathbf{F}_p$ strongly departs from the isotropic case and for $\theta=\theta_c$ coincides with the direction of the \v{C}erenkov cone given by eq.~(\ref{cerenkov_sym_ext_crit}). This follows directly from eq.~(\ref{forcesolution}). For $\theta_c$ the waves are emitted only into one direction (the \v{C}erenkov cone becomes a line); therefore $k_x$ and $k_y$ are proportional and the ratio $F_y/F_x$ is the same as $k_y/k_x$. The longitudinal force is thus parallel to the symmetry axis of the \v{C}erenkov cone. For $\theta > \theta_c$ the direction of $\mathbf{F}_p$ departs from that of the cone.

The transverse force $F_z$ is a result of a complex interplay between evanescent and \v{C}erenkov interactions (see~\cite{schieber:1998, ribic} for an explanation of the isotropic case); nevertheless, $F_z$ only slightly varies with $\theta$ (within 10\% in the interval $\theta \in [0, 90^{\circ}]$). As in the isotropic case~\cite{ribic}, $F_z$ is always attractive (negative) for a point charge; it cannot be made repulsive simply by increasing $\beta$ or changing the dielectric constant. However, it can become repulsive by replacing the point charge with a transverse line of charge.

For $\beta=0.5$ the \v{C}erenkov condition for extraordinary waves is satisfied for all $\theta$. For low angles $\psi$ is above the value for the isotropic case: $\psi > \theta$. At some critical angle we enter the regime $\psi<\theta$. Increasing $\theta$ leads to transition back to the regime $\psi>\theta$. This peculiar behavior cannot be qualitatively explained by considering the direction of the \v{C}erenkov cone; the direction of the force has to be determined by integration over all the momenta of the waves excited in the solid.
\begin{figure}
\includegraphics[width=0.35\textwidth]{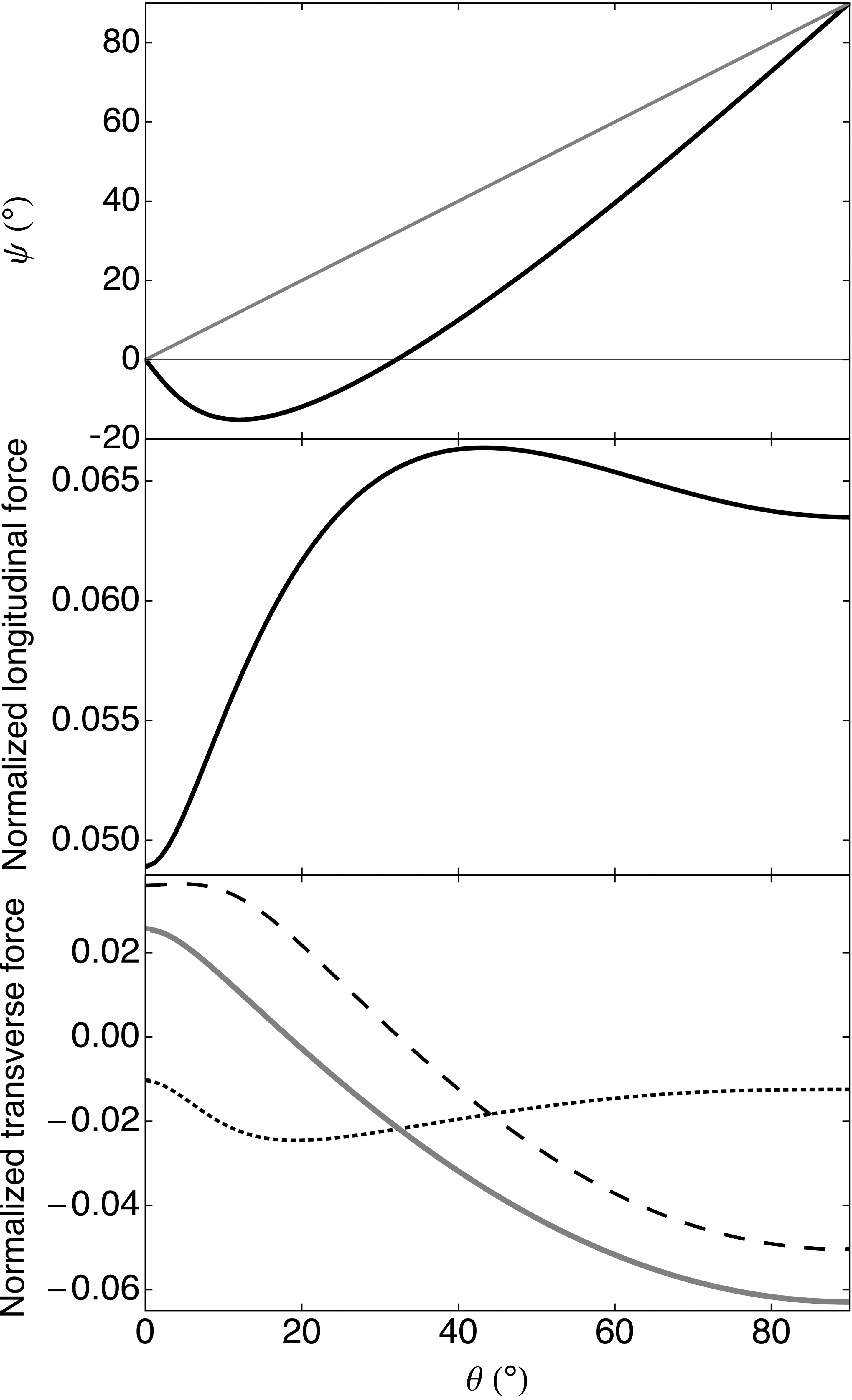}
\caption{$\psi=\arctan{(F_{y}/F_{x})}$ (top panel), magnitude of the longitudinal force (middle panel), and transverse force (bottom panel) as a function of $\theta$ for a plasma type response with $\omega_{p1}=0.25\omega_{max}$, $\omega_{p2}=10^{-1}\omega_{p1}$. Solid black line in the top panel represents the calculated values, dashed line is the isotropic case. In the bottom panel the transverse force (thick gray line) is split into the \v{C}erenkov (dashed) and evanescent (dotted) contributions.}
\label{fig.5}
\end{figure}

For higher $\beta$ the \v{Cerenkov} condition for ordinary waves is fulfilled. These waves contribute to a force in the direction opposite to the particle velocity. The direction of $\mathbf{F}_p$ therefore approaches that of the isotropic case, as demonstrated in the bottom panel of fig.~\ref{fig.3}.

We note that for the two extreme points $\theta = 0^{\circ}$ and $\theta = 90^{\circ}$ the longitudinal force always points in the direction opposite to $\mathbf{v}$, which follows from the symmetry of the situation.

As $\beta$ increases the particle excites more \v{C}erenkov waves and therefore the magnitude of $\mathbf{F}_p$ increases, as indicated in the right column of fig.~\ref{fig.3}. For $\beta=0.45$ the maximum value of $F_p$ is $\sim2$ orders of magnitude smaller than $F_z$; however, the components become comparable at $\beta=0.7$.

For the reversed situation, $\epsilon_1 = 4$ and $\epsilon_2 = 8$, the anisotropy is much less pronounced since with increasing $\beta$ ordinary \v{C}erenkov waves are excited first and these contribute to a force in the direction opposite to $\mathbf{v}$. The effect (not shown here) is similar to the situation in the bottom panel of fig.~\ref{fig.3}, left column, except that $\psi < \theta$.

Figures~\ref{fig.4} and~\ref{fig.5} show the results for a plasma type dielectric: $\gamma_{1,2}=10^{-2}\min({\omega_{p1},\omega_{p2},\omega_{max}})$. We had to work with finite losses to insure numerical convergence. Although the conditions for a plasma type dielectric are violated at $\omega=0$ and near $\omega_{p1,p2}$, insuring that $\omega_{max} \gg \gamma_{1,2}$ and putting the plasma frequencies "far" apart (e.g., $\omega_{p2}=10\omega_{p1}$), the contribution to the integral around these points is small (the numerical results are nevertheless exact because we are using the Drude model for integration).

Figure~\ref{fig.4} demonstrates that the force direction as well as the magnitude change with $\omega_{max}$. Depending on $\omega_{max}$ the angle $\psi < \theta$ or $\psi > \theta$. Since $\omega_{max}$ depends on the particle speed $v$ and the distance $z_0$ from the dielectric, the force magnitude as well as the force direction also depend on $v$ and $z_0$. This is in strong contrast to the case of a dielectric with no losses and no dispersion, where the force direction is independent of $z_0$. 

Interesting behavior is observed when $\omega_{max}>\omega_{p1}>\omega_{p2}$,  fig.~\ref{fig.5}. The angle $\psi$ is negative for low $\theta$, which means that the $y$ components of the force and velocity have the same sign and the particle is accelerated in the $y$ direction. Nevertheless, this does not violate energy conservation because the product $\mathbf{v} \cdot \mathbf{F}_p <0$ and the particle's energy decreases.

Peculiar behavior of the transverse force is observed for values of $\theta$ below $\approx 20^{\circ}$ where it becomes repulsive (positive). This is shown in the bottom panel of fig.~\ref{fig.5}, where we split the interaction into the \v{C}erenkov part ($\omega_{p2} < \omega < \omega_{p1}$) and the evanescent part ($0 < \omega < \omega_{p2}$ and $\omega > \omega_{p1}$). For this particular case the \v{C}erenkov contribution, which can become repulsive, starts to dominate for low angles. The origin of the repulsion is the momentum carried into the dielectric by the excited waves, balanced by the particle ~\cite{schieber:1998}. It is therefore possible for a dielectric surface not only to repel a transverse charge packet as demonstrated in ref.~\cite{ribic}, but also a point charge; an outcome which seems to contradict common notions based on electrostatic considerations of point charges above metallic or dielectric surfaces.

\section{acknowledgments}
The research was in part supported by the Fonds National Suisse (FNS) de la Recherche Scientifique and by the CIBM.

\end{document}